# Network Motifs in Object-Oriented Software Systems

Yutao Ma, *Member*, ACM
State Key Lab of Software Engineering
Wuhan University
Wuhan, China
yutaom@acm.org

Keqing He, *Senior Member*, *IEEE*
State Key Lab of Software Engineering
Wuhan University
Wuhan, China
hekeqing@public.wh.hb.cn

Jing Liu
State Key Lab of Software Engineering
Wuhan University
Wuhan, China
jingliu@sklse.org

*Abstract*—Nowadays, software has become a complex piece of work that may be beyond our control. Understanding how software evolves over time plays an important role in controlling software development processes. Recently, a few researchers found the quantitative evidence of structural duplication in software systems or web applications, which is similar to the evolutionary trend found in biological systems. To investigate the principles or rules of software evolution, we introduce the relevant theories and methods of complex networks into structural evolution and change of software systems. According to the results of our experiment on network motifs, we find that the stability of a motif shows positive correlation with its abundance and a motif with high Z score tends to have stable structure. These findings imply that the evolution of software systems is based on functional cloning as well as structural duplication and tends to be structurally stable. So, the work presented in this paper will be useful for the analysis of structural changes of software systems in reverse engineering.

## I. Introduction

Nowadays, complex networks are studied across many fields of science and engineering. Many of them that occur in both the organic forms of nature and the engineered artifacts of human society have been shown to share global statistical features such as "Small World" property [1] (this phenomenon is also popularly known by "six degrees of separation" [2]), and "Scale Free" property [3] (if a node has $k$ edges, the degree distribution $P(k)$ of the network decays as a power law $P(k) \sim k^{-r}$, where $r$ is often between 2 and 3).

Furthermore, many class diagrams of large-scale object-oriented (OO) software systems have also been found to share "Small World" and "Scale Free" properties [4,5,6]. Hence, Myers thinks that large-scale software systems represent an important class of artificial complex networks [4]. This unexpected result raises theoretical questions about the traditional principles by which software systems form and evolve. Software systems, no matter how they are designed, are subject to continuous evolution and maintenance activities in order to eliminate programming errors and defects and to extend their functionalities [7]. So, researchers from different research domains are trying to find out and explain how software systems evolve and remain robust and adaptable in the face of changing environments [8].

In general, the analysis of system structure plays a fundamental role in exploring principles of software evolution [9]. Without explicit and immediate support for structural evolution, software systems may become unnecessarily complex and unreadable, which results in many emergent problems as they are adapted to changing requirements [10]. To discover the principles of software evolution would require an understanding of the basic structural elements particular to each class of networks. So, network motifs, patterns of interconnection occurring in complex networks at numbers that are significantly higher than those in randomized networks, are defined to uncover structural design principles of networks [11]. They are suggested to be elementary building blocks that carry out key functions in the network.

Software developers often build different software systems with the same internal library, framework, or pattern that is typically applied at a small scale. These components describing interactions among 3 or 4 classes or objects are similar to network motifs in terms of their micro structures. Hence, in virtue of network motifs, it is interesting to explore how micro-structural design methodologies conspire in the large to form macroscopic software structures, which is still a challenging issue in software engineering [4]. In this paper, our work focuses mainly on the problem that what property network motifs possess may drive the structural evolution of OO software systems. We argue that this will provide a new insight into the structural evolution of OO software systems, which may facilitate understanding how software systems evolve over time. Based on the work presented in this paper, we hope to be able to give developers information about worthwhile evolutionary trends in OO software systems that they are working on.

This paper is arranged as follows: Section 2 introduces the related work; in Section 3, we use a free tool to analyze the common motifs in six OO software systems, and find that they possess some important properties and motifs with high Z scores tend to have stable structures; in the end, Section 4 concludes the paper and puts forward the future work.

## II. RELATED WORK

### A. Software Engineering

Recently, Fioravanti et al. analyzed structural similarities in C++ software at the module (class) level and found quantitative evidence of structural duplications [9]. Clones of function and structure in software systems or web applications [12] are similar to the evolutionary trends (e.g. self-reproduction) found in biological systems. However, they did not provide any model to explain the origin of duplications [13]. On the other hand, the current mainstream research work of reverse engineering pays close attention to architectural recovery, analysis of class diagrams derived from source codes, program slicing, and so on. In general, they were performed at the lower level such as source lines of code (LOC) and class, which results in insufficient abilities to describe and measure the overall structure of large-scale software systems [14].

### B. Complex Networks

Motifs in a network are small connected sub-graphs that occur in significantly higher frequencies than in randomized networks. They have recently attracted much attention as a useful concept to uncover structural design principles of complex networks. Valverde et al. analyzed a large set of software class diagrams and found that dynamical rules, with little relation to underlying functional constraints, largely determine the frequency of motifs in software graphs [13]. Because they focused on the duplication and growth of sub-graphs (motifs) in software systems, their work is different from the traditional studies of software evolution that lay emphasis on the dynamics of an individual object or a whole class diagram. However, they failed to explain the mechanism of dynamical rules in their growth model in terms of the relevant domain knowledge of software engineering, which leaves a lot of details unexplained.

## III. STABILITY-DRIVEN STRUCTURAL EVOLUTION

It is useful to depict complex structure defined in software programs by means of a directed graph or a network model, where nodes represent software entities (e.g. class, module, etc.) and links represent different kinds of relationships between classes, modules, and instructions (e.g. inheritance, call, etc.). For details on the method for transforming a software system into a directed graph, please refer to [15].

### A. Tool and Software Systems

Three software collaboration graphs (VTK, Digital Material, and AbiWord) available at http://www.tc.cornell.edu/~myers/Data/SoftwareGraphs, two open-source software systems (Tomcat 5.0 and loki 0.1.2) available at http://sourceforge.net, and a commercial Java system (SCRR) [16] developed by us will be examined in this paper. The tool used to analyze network motifs (mfinder 1.2) is available at http://www.weizmann.ac.il/mcb/UriAlon/. Considering the common size of micro-structures in software systems, we will mainly investigate the occurrence and significance of network motifs that have 3 or 4 nodes in six different software systems.

### B. Data Analysis

The statistics of subgraphs provides important information about network structure. The result of our analysis on 6 software networks is shown in Table I. For each motif, we list the number of occurrences in the real system ($N_{real}$), the number of occurrences ($N_{rand} \pm SD$) in a set of 100 randomized networks, and a qualitative measure of its statistical significance as given by the Z score [11]. $SD$ denotes the standard deviation. All parameters of mfinder 1.2 are set by default.

A handful of motifs appear to be present in all software systems analyzed. Such a common phenomenon may imply that similar subgraphs are abundant because they are selected or chosen to perform a given function or task. However, semantic ambiguity in the functional meaning of motifs (without OO contents and contexts) suggests that motifs in software networks are not strictly related to well-defined functions, but possibly reveal the structural constructs of software systems.

TABLE I. COMMON MOTIFS IN ALL SIX SOFTWARE SYSTEMS

| System | Nodes | Edges | $N_{real}$ | $N_{rand}$ | $Z_{socre}$ | $N_{real}$ | $N_{rand}$ | $Z_{socre}$ | $N_{real}$ | $N_{rand}$ | $Z_{socre}$ |
|---|---|---|---|---|---|---|---|---|---|---|---|
| Motif ID | | | | 38 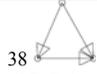 | | | 204 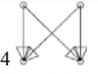 | | | | |
| loki | 122 | 138 | 31 | 9.8±3.0 | 7.07 | 42 | 18.5±6.3 | 3.73 | | | |
| DM | 187 | 271 | 49 | 19.0±4.3 | 6.91 | 82 | 30.8±9.5 | 5.40 | | | |
| Motif ID | | | | 38 | | | 204 | | | 344 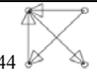 | |
| SCRR | 598 | 1209 | 107 | 56.3±8.4 | 6.01 | 2927 | 750.1±54.2 | 40.19 | 33 | 9.5±3.5 | 3.01 |
| AbiWord | 733 | 739 | 21 | 0.9±1.1 | 18.66 | 7 | 0.5±0.7 | 9.94 | 15 | 1.1±1.8 | 7.69 |
| VTK | 788 | 1374 | 229 | 77.4±9.5 | 15.96 | 1436 | 590.8±97.1 | 8.70 | 393 | 113.3±23.1 | 12.13 |
| Tomcat | 1751 | 1757 | 276 | 74.8±12.3 | 16.30 | 512 | 246.5±39.8 | 6.67 | 245 | 52.7±12.4 | 15.56 |
| Motif ID | | | | 904 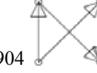 | | | 2186 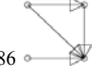 | | | 206 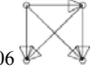 | |
| SCRR | 598 | 1209 | 25 | 10.6±4.6 | 3.15 | 30 | 13±6.7 | 2.53 | 21 | 1.5±1.3 | 14.80 |
| AbiWord | 733 | 739 | 6 | 0.3±0.6 | 10.03 | 10 | 1.9±1.7 | 4.66 | | N/A | |
| VTK | 788 | 1374 | 295 | 65.5±13.2 | 17.32 | 5828 | 2983.8±541.1 | 5.26 | 41 | 6.5±3.7 | 9.22 |
| Tomcat | 1751 | 1757 | 65 | 34.6±11.3 | 2.70 | 10212 | 5559.7±1148.6 | 4.05 | 127 | 7.8±4.3 | 27.46 |

Previous studies have presented the idea that network motifs seem to define the minimal, meaningful building blocks of complex networks [11]. However, some subgraphs appear not to be common motifs that recur in different software systems only according to Z score (in general, when Z > 2 the motif is considered to be more common than expected from random networks). In Figure 1 we present the distributions of all 3- and 4-node motifs in our chosen software systems to show their real occurrences. X axis represents the motif ID (for 4-node motifs, there are total 199 enumerations, so motif's ID labeled 1 corresponds to the real ID "14", and so on), and Y axis indicates the occurrence or abundance of a motif (log $N_{\text{real}}$) in real systems. Note that ID labels match the output from mfinder 1.2.

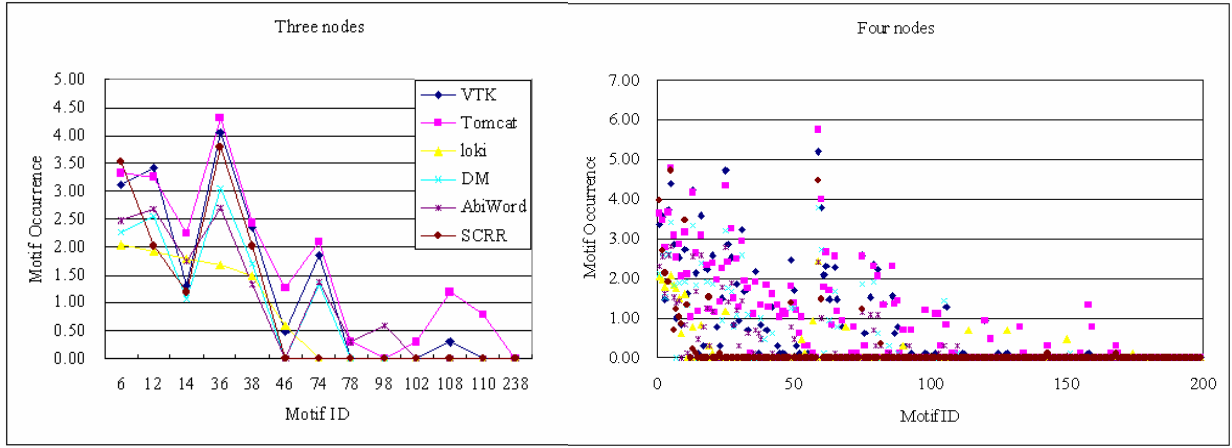

Figure 1. Occurrence of motifs with 3 or 4 nodes

According to the figure presented here, we can find that the occurrence tends largely to decrease along with the increase of a motif's ID number, i.e., motifs with high frequencies are sparser (fewer links) than those subgraphs, which are more dense. For example, in the left part of Figure 1, motifs (e.g., 12 and 36) comprising two edges might be encountered more frequently than motifs with a higher number of edges such as 74. This suggests a very simple duplication-based mechanism of subgraph generation, which implies that simple structure and fewer constraints on interactions among its components may enable a micro-structural subgraph to achieve easier reproduction. For software engineering, this is also a recognized law, namely, a simple structure comprising fewer constraints may foster broader reusability. We think that this provides the statistical evidence to prove the structural reasonableness of motifs with high occurrence. However, what property may determine the non-random structure of these motifs?

*C. Measurement of Structural Stability*

Recently, Prill et al. discovered that stability or robustness to small perturbations is highly correlated with the relative abundance of small network motifs in several previously determined biological networks [17]. Traditionally, the structural stability of an OO subsystem is a sign of its capability to evolve while preserving its structure [18], which is similar to the robustness defined in system engineering.

As we know, the Jacobian matrix can be used to denote the local connectivity of a motif if the term $a_{ij}$ represents the sign and weight of influence of the $j^{\text{th}}$ node onto the $i^{\text{th}}$ node. Then, we make use of the eigenvalues of a concrete matrix to determine whether the system will become stable. Based on the method in [19], Prill et al. defined a metric (Structural Stability Score, SSS) to indicate the probability that the dynamical system corresponding to a given motif relaxes monotonically to steady-state following a small perturbation [17]. For example, we find that these motifs (SSS = 1) are directed acyclic subgraphs devoid of feedback loops. In general, the stability is associated with the gain of a feedback loop; when the number of loops or the size of a loop increases, the stability decreases. This is similar to the popularly-known McCabe cycle [20] in software systems. As the number of McCabe cycles increases, the complexity increases and then the stability would decrease. The second class of motifs (SSS ≈ 0.5) contains a single 2-node feedback loop. If you pick signs of edges of the loop from a uniform distribution, then 1/2 the time the loop will have a negative gain, which certainly makes the motif structurally stable. The third class of motifs (SSS < 0.2) contains more complicated circuits: multiple 2-node loops, 3- or 4-node loops, etc. Their stability can't be guaranteed by specifying the sign of the feedback loops present.

According to the classification, 3-node motifs will be categorized into 3 groups: 1) 6, 12, 36, and 38; 2) 14, 46, 74, and 108; and 3) 78, 98,102,110, and 238. It is visually apparent in Figure 1 (left part) that the occurrence is correlated with the stability class, i.e., structurally stable motifs often have higher occurrence than those with lower stability class. Moreover, for motifs with the same stability class, the occurrence is correlated with the number of edges of the motif, e.g., motif 38 (with 3 edges) has less overrepresentation than those with 2 edges (such as motif 6, 12, and 36). Although 4-node network motifs capture a richer representation of the local connectivity patterns than the 3-node profiles, the general trend of structural stability is the same as in the 3-node analysis. This implies that the reusability (duplication) may be driven by the requirement of structural stability to small perturbations, such as internal failures or the evolvability.

## D. Relationship between Z Score and Stability

Remarkably, in all systems analyzed, the stability class shows excellent correlation with the occurrence. Nevertheless, whether these common motifs with high Z score have stable structures? This is an interesting problem, which will facilitate the understanding of structural evolution in software systems. Above all, the Z score profiles are normalized to unit vectors to enable comparison of scores across different systems, and the normalized Z score [21] is defined as $N_{z_i} = Z_i / \sqrt{\sum Z_i^2}$ .

All 3-node motifs are sorted on the X-axis first, according to increasing number of edges. Normalized Z scores (blue bars) for 13 kinds of motifs in the chosen software systems are shown with outlines of stability class (red outline provided as a guide to the eye [17]). Each black dashed line indicates a change in the number of motif edges, viz. the boundary of specific groups based on the number of edges. According to the 3-node profiles presented here (see the left part of Figure 2), motifs with high Z score tend to have stable structure; moreover, motifs with highest Z score also have more stable structure than the other motifs with the same number of edges (see the right part of Figure 2), for example, motif 38 (class I) and 108 (class II) in Tomcat has higher Z score than the other motifs with 3 edges (such as motif 14 and 98) and 4 edges (such as motif 78), respectively. Note that although some motifs (class III) have relatively high Z score, they seldom recur in real systems, so we ignore them when analyzing the relationship between Z score and stability class in the 3- and 4-node profiles.

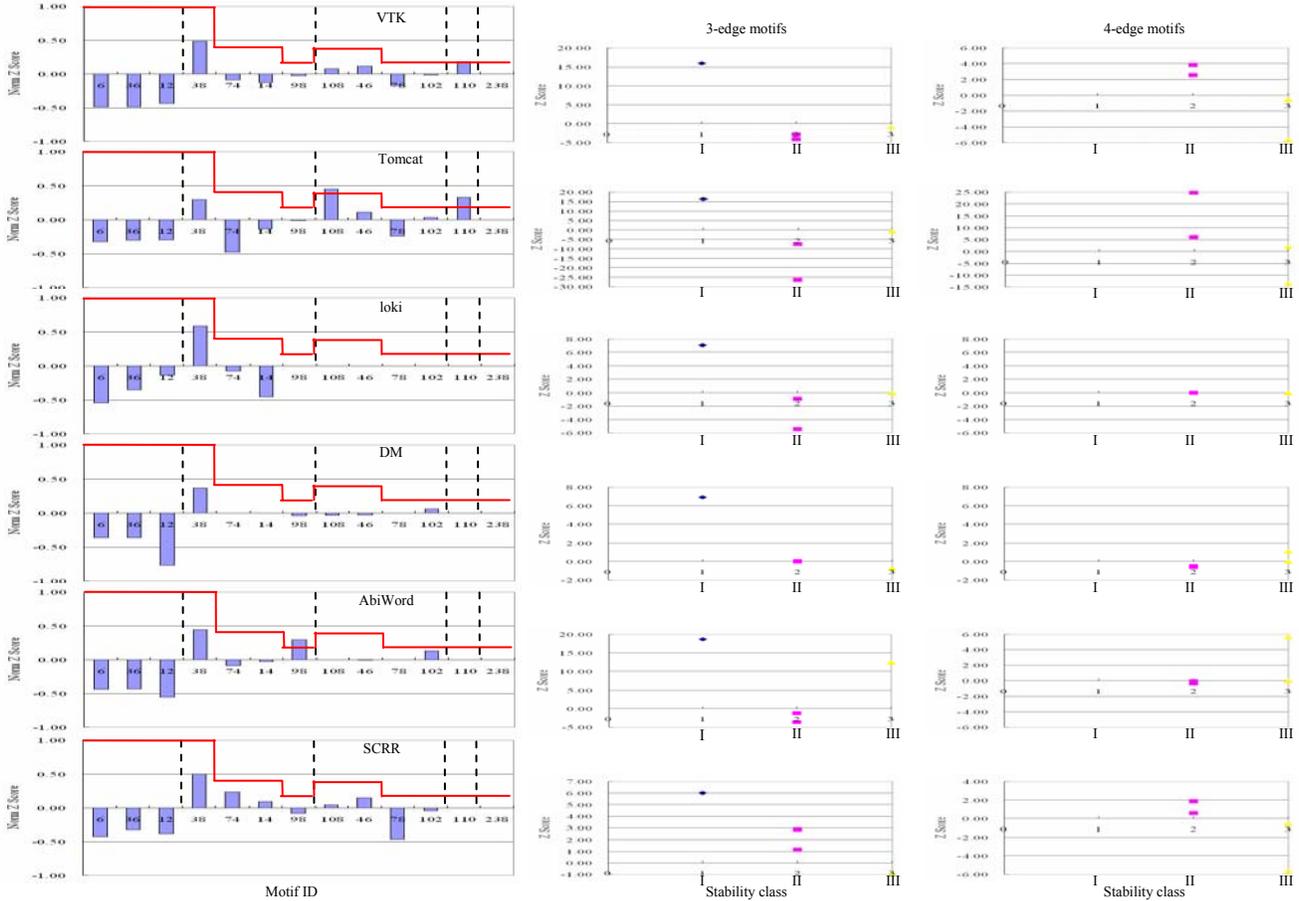

Figure 2. Distribution of normalized Z scores of 3-node motifs

To prove the broad validity of our finding, we present the distribution of normalized Z scores of 4-node motifs in Figure 3. All 199 kinds of motifs are sorted on the X-axis first, according to increasing number of edges. The other graphical notations are the same as those defined in Figure 2 (left part). In Figure 4 we show Z scores of 4-node motifs classified by the stability class within specific groups based on the number of edges (4, 5, and 6). These groups contain at least one motif in each stability class. Box and whisker plots mean a difference in the average Z score between stability classes [17]. The plot is interpreted as follows: the box indicates the inner (the first and the third) quartile range; the average Z score is denoted by a horizontal red line; the whiskers extend to cover the upper and lower quartiles up to a distance of one time the inner quartile range; red dots mean the extreme scores (beyond the whisker's length); $p$-values attached to each plot express the probability that the observed difference in Z score between stability classes is expected by chance. We calculate them by using the Kruskal-Wallis test function of Analyze-it 1.73 (a trial version available at www.analyze-it.com). In most cases, the difference in Z score between stability classes is significant at the usual criterion of 95% confidence or better.

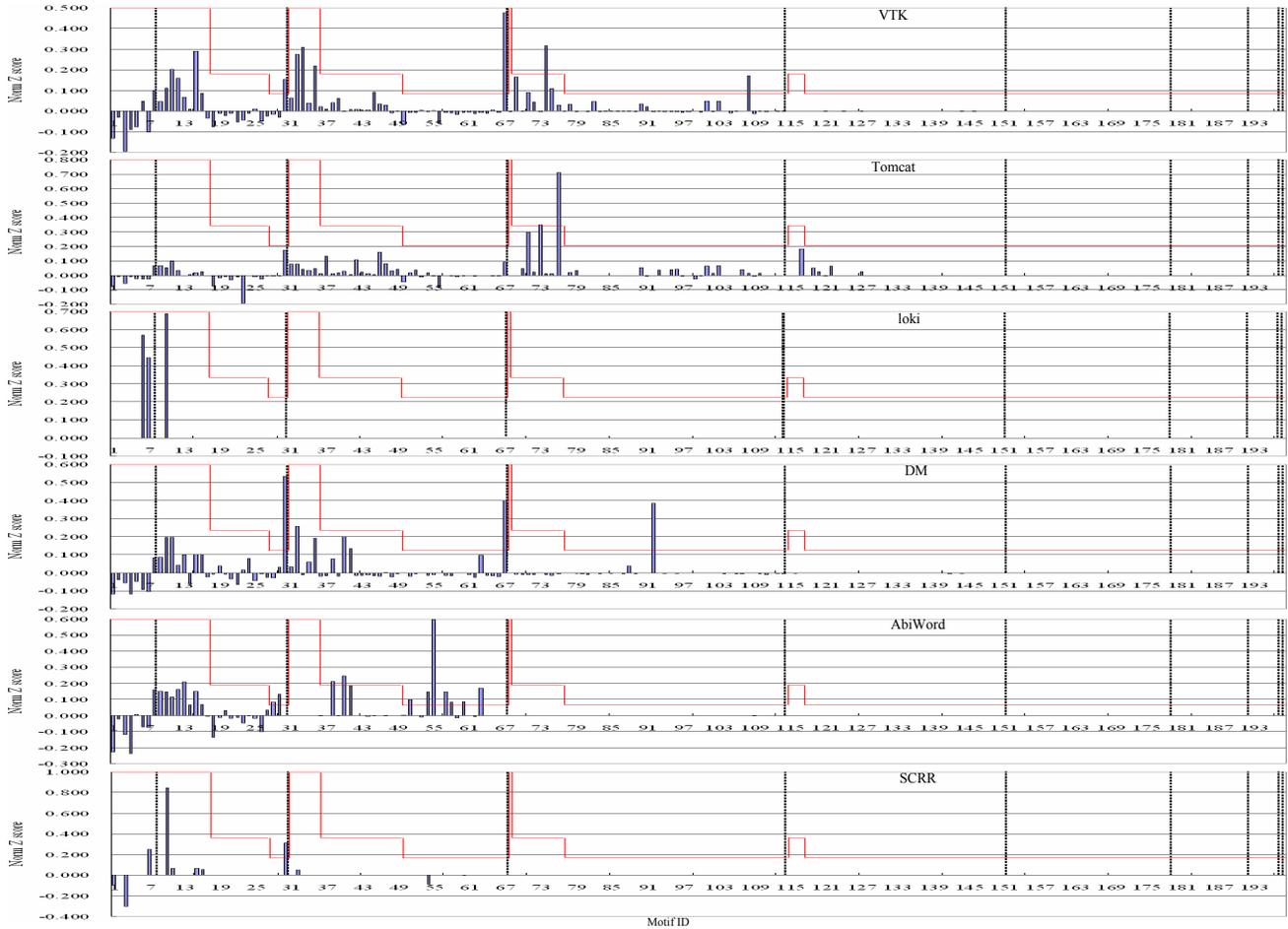

Figure 3. Distribution of normalized Z scores of 4-node motifs

As we expected, the general trend of structural stability is the same as in the 3-node analysis. The systems differ in precisely which motifs are overrepresented (see Figure 3), but the dynamic properties of overrepresented motifs (namely, motifs with high Z score often have stable structure) are conserved across all systems presented here. Furthermore, we also find that there is similarity in stability properties among highly overrepresented motifs. We select the most significant motifs in each specific group across all chosen systems by setting a filter condition (Z score > 2 and Mfactor > 1.1 and Uniqueness >= 4, see the manual of mfinder 1.2), and discover that they (most are presented in Table I) belong to class I or II and dominate the non-random organization of the network within their specific groups. So, this demonstrates the preference for structural stability among the highly overrepresented motifs within the specific groups. For example, the 5-edge specific group (between the second and third black dashed line) of 4-node motifs is comprised of 37 motifs. Among these motifs, 6, 14, and 17 motifs belong to class I, II, and III, respectively. VTK contains a highly overrepresented motif of class I (2190, labeled as 34 in Figure 3), which has the highest Z score within the specific group. The remaining relatively high Z scores correspond to the motifs of class II, and none of the 17 motifs with low stability have high Z score.

In addition to statistically significant differences in average Z score among stability classes (see Figure 4), the similarity among all examined systems is the overrepresentation of stable motifs compared to the other motifs with the same number of edges (see the left part of Figure 2 and Figure 3). In general, motifs with high Z score tend to have stable structure, but some motifs that have the most stable structure (belong to class I) don't always show high Z scores (see the group before the first black dashed line in the left part of Figure 2 and in Figure 3). This implies that structural stability may be necessary, but not sufficient, for network motif overrepresentation [17]. Even so, we argue that these findings in OO software systems unfold a fundamental rule (viz. stability-driven duplication or cloning) of structural evolution, which is different from what found in random networks (because of lacking any organizing principles, the distribution of motifs in random network such as Erdös-Renyi graph is determined by the density of edges [22]).

E. *Implications for Software Engineering*

Scale-free networks are more stable or robust to random attacks than simple random networks [8]. From the top-down view of network decomposition [23], the global feature of scale-free networks should be analyzed at multiple levels—from the graph-theoretic view, through hierarchical levels of subsystems, and down to individual network motifs. So, network motifs are deemed as a kind of variable that distinguishes real-world networks and random networks. Based on our findings in OO software systems, we believe that the robustness of scale-free software systems may stem from the structural stability of their basic building blocks (viz. network motifs) under constrained growth in the process of software evolution. However, the origin of this constrained growth remains to be explained.

It is widely acknowledged that software is probably one of the most intricate human inventions. For software development, it is important to follow the recognized principles: (1) efficient communications among software entities at low cost, that is, a software system should have a relatively small average shortest path length; (2) high cohesion among components within an entity and low coupling between entities (simply, High Cohesion and Low Coupling), which explains why software systems always have larger clustering coefficient by means of the encapsulation of OO technique (namely, simpler related functions are grouped into entities to achieve high cohesion); and (3) information transfer or interaction should keep working in an efficient way when a randomly chosen class fails, that is, since a system's sensitivity to component failure is a fundamental problem in any area of engineering, the structure of software systems should be reliable or stable, otherwise, it will influence functions and performance of a system. The above principles of software engineering can provide an insight into why software systems have "Small World" and "Scale Free" features.

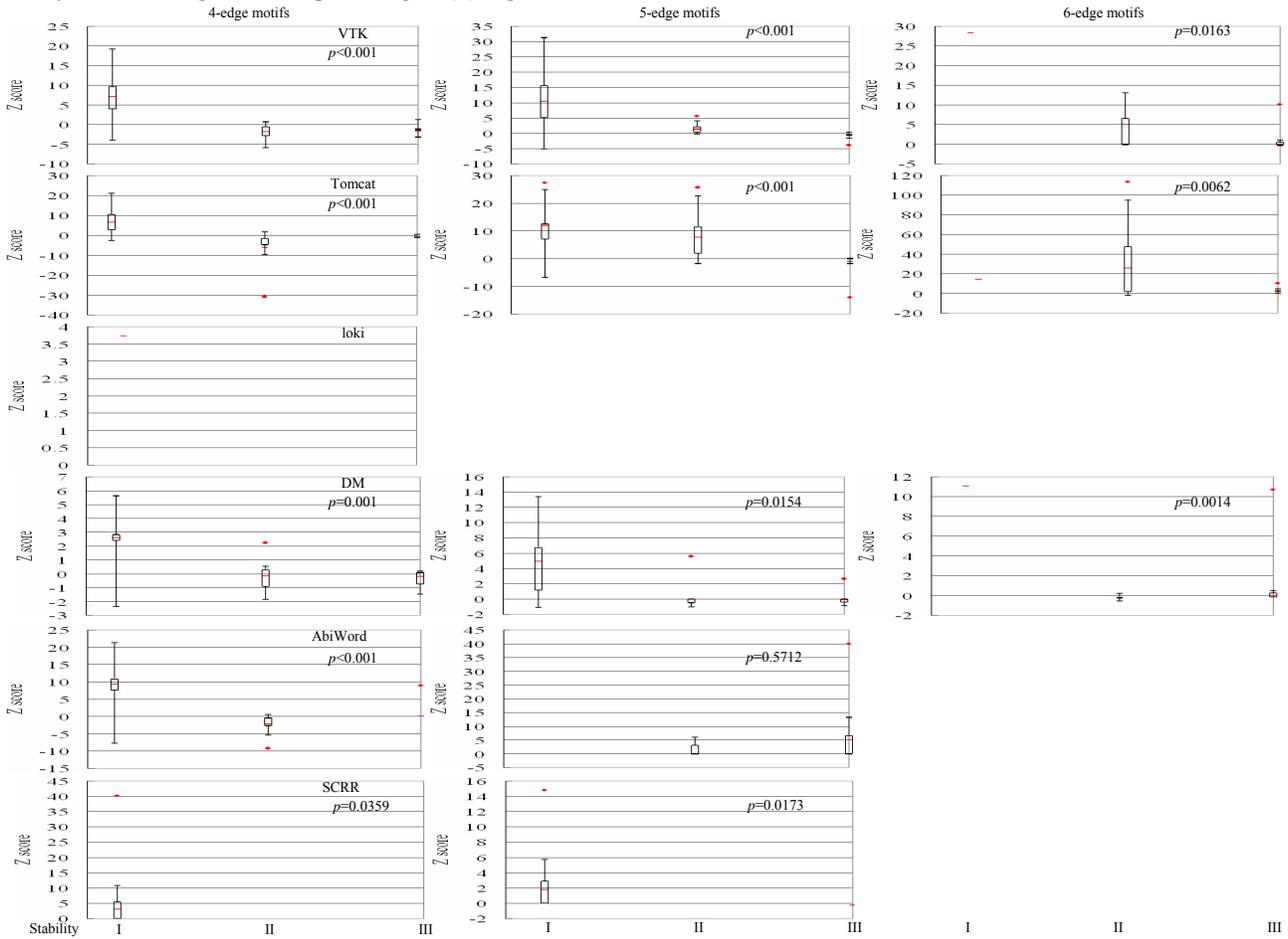

Figure 4.  Z score of 4-node motifs classified by the stability class within specific groups

Furthermore, in the process of software development (note that we leave some human factors such as budget, programmer's skills, and external pressures out of consideration), the structure of software systems tends to evolve from disorder to order in terms of structural entropy [15]. Eventually, they will evolve to scale-free networks under some universal principles of software development. So, the "Scale Free" feature of software systems is an emergent property of software evolution [24]. If so, we argue that software systems always tend to become more stable or robust and to form ordered structure during their evolution in order to perform pre-designed functions well. In order to achieve small average shortest path length and stable structure, software systems evolve over the time of development process and then form scale-free networks, indicating the general trend of software evolution, especially structural evolution. Hence, we think the evolution due to global constraints on network structure (such as structural stability or robustness) can create network motifs, which accords with Valverde's conclusions [13]. On the other hand, network motifs with high Z score in scale-free software networks also have been found to possess stable structure, possibly explaining a recognized design principle that loops should be avoided when software engineers design their programs [25]. This is an interesting by-product in our paper.

IV. CONCLUSIONS AND FUTURE WORK

Now, more and more researchers realize that large-scale software systems represent an important class of artificial complex networks, which possess global statistical features such as "Small World" and "Scale Free". They utilized different approaches to uncover the evolutionary rules of software systems under a changing environment. Our work is inspired by the previous work that detected and analyzed motifs in networks from biochemistry, neurobiology, and OO software. The micro-structural duplication based on network motifs may provide a new insight into the structural evolution. According to the results of our experiment, motifs with stable structure comprising few constraints often have higher occurrences in real systems, possibly implying that robustness or stability enables a motif to achieve easier and broader reusability (duplication); motifs with high Z score tend to have stable structure, and these overrepresented motifs are significant ones that act as basic building blocks of complex systems, so we believe that structural evolution of software systems is based on stability-driven functional cloning and structural duplication. Then, the work presented in this paper will facilitate our understandings of software evolution.

Evolution of industrial quality software systems is notoriously expensive, so it is therefore paramount to investigate the flexibility or evolvability of software and to find ways to quantify it. So, the future work is to investigate the evolution complexity of micro-structures in terms of network motifs. Based on the relevant metrics, we can quantify the costs of change in software structure and measure the influences on the whole structure.

ACKNOWLEDGMENT

This work is supported by National Basic Research Program (973) of China under grant No. 2006CB708302, National High Technology Research and Development Program (863) of China under grant No. 2006AA04Z156, National Natural Science Foundation of China under grant No. 90604005, Provincial Natural Science Foundation of Hubei Province of China under grant No. 2005ABA123 and No. 2005ABA240, and the Open Research Foundation of State Key Laboratory of Software Engineering of China under grant No. SKLSE 05-03.

We gratefully acknowledge the generous help of Dr. R. J. Prill (Johns Hopkins University), Professor B. Li, and Dr. D. Du (SKLSE).